# Identifying Data Noises, User Biases, and System Errors in Geo-tagged Twitter Messages (Tweets)

**Ming-Hsiang Tsou**
Center for Human Dynamics in the Mobile Age, San Diego State University, 5500 Campanile Drive, San Diego, CA 92182, USA.
01-6195940205
mtsou@mail.sdsu.edu

**Hao Zhang**
Center for Human Dynamics in the Mobile Age, San Diego State University, 5500 Campanile Drive, San Diego, CA 92182, USA.
zhanghaoshogo@gmail.com

**Chin-Te Jung**
Esri (Beijing) R&D Center, Beijing, China
chinte.jung@gmail.com

## ABSTRACT
Many social media researchers and data scientists collected geo-tagged tweets to conduct spatial analysis or identify spatiotemporal patterns of filtered messages for specific topics or events. This paper provides a systematic view to illustrate the characteristics (data noises, user biases, and system errors) of geo-tagged tweets from the Twitter Streaming API. First, we found that a small percentage (1%) of active Twitter users can create a large portion (16%) of geo-tagged tweets. Second, there is a significant amount (57.3%) of geo-tagged tweets located outside the Twitter Streaming API's bounding box in San Diego. Third, we can detect spam, bot, cyborg tweets (data noises) by examining the "source" metadata field. The portion of data noises in geo-tagged tweets is significant (29.42% in San Diego, CA and 53.47% in Columbus, OH) in our case study. Finally, the majority of geo-tagged tweets are not created by the generic Twitter apps in Android or iPhone devices, but by other platforms, such as Instagram and Foursquare. We recommend a multi-step procedure to remove these noises for the future research projects utilizing geo-tagged tweets.

## 1. INTRODUCTION
Twitter is one of the most popular social media platforms used in academic research works due to its large number of users, comprehensive metadata, openness of messages, and public available application programming interfaces (APIs). Tweets are the actual Twitter messages created by users to express their feelings, events, or activities within 140 characters including spaces. Geocoding tasks for tweets are important for social media analytics because data scientists and researchers need to aggregate social media messages or users into a city, a region, or nearby points of interests (POI) for location-based analysis and regional trend analysis. Currently, the public Twitter Application Programming Interfaces (APIs) can provide five types of geocoding sources: 1. Geo-tagged coordinates, 2. Place check-in location (bounding box), 3. User Profile Location, 4. Time Zones, 5. Texts containing locational information (explicit or implicit information). Among the five geocoding methods, tweets with geo-tagged coordinates is the most popular data source used in location-based social media research. Geo-tagged tweets have precise latitude and longitude coordinates (decimal degrees) stored in a metadata field of tweets, called "geo" (a deprecated field name in APIs) or "coordinates" (the current field name in APIs). When users turn on the precise location tag function on their Twitter accounts (which is off by default), their tweets will be geo-tagged using GPS or Wi-Fi signals in their mobile devices. Since many users do not enable precise location tags, there are only around 1% of tweets containing geo-tagged information. The percentage of geo-tagged tweets may vary among different topics or keywords. For example, during a wildfire event, the percentage of geo-tagged tweets can become 4% or higher collected by using wildfire related keywords [1].

Geo-tagged tweets are valuable data for social media researchers and geographers to study geographic context and spatial association within social media data. One popular method of collecting geo-tagged tweets is to utilize Twitter's Streaming API with a predefined bounding box or multiple predefined keywords. Previous works have demonstrated that social media data collected by the Twitter Streaming API (free version with 1% sampling rate) are a good sample of Twitter's Firehose API (very expensive version providing 100% tweet data in Twitter servers) [2]. In academics, many researchers used geo-tagged tweets for conducting spatial analysis and GIS operations for their research projects. For example, the 2013 special issue of "*Mapping Cyberspace and Social Media*" in Cartography and Geographic Information Science [3] includes seven refereed research papers and four out of seven paper are using geo-tagged tweets as their main data sources. There are two types of Twitter APIs in general, Streaming API for collecting real-time feeds of Twitter messages and Search API for collecting historical tweets (up to 7 or 9 days before the search date) with specific keywords or user names with database query methods. This paper will only focus on the characteristics of geo-tagged tweets collected from the Twitter Streaming API.

## 2. TWITTER SPAMS, BOTS, AND CYBORGS
Previous research has identified three major types of data noises in Twitter data, spams, bots, and cyborgs. Twitter spams, similar to email spams, vary in different forms. Some spams are transparent and easy to be identified while some spams are sophisticated [4]. Spams are usually created for reaching more users and increase the financial gain for spammers. Spammers now use multiple platforms to disseminate spams which also include social media services [5]. Many researchers have addressed the importance of eliminating spams for a clearer and safer online networking environment [6]. The existence of spams in Twitter was first noticed in 2008 by studying over 100,000 Twitter users and grouping them base on follower-to-following ratios [7]. The groups contain broadcasters (users with large amount of followers), acquaintances (users who have a follower-to-following ratio close to 1), and miscreants (users who follow a lot and less followed) [7]. Many Twitter spams include URLs. According to Grier et al. [8], 8% of the URLs shared on Twitter are considered as the linkage to malicious websites, which contains malware or scams [8]. Clicking on those embedded

URL may cause serious problems to users' privacy and local computers [9]. The harmfulness of spams does not only limit to malware and scams. Spam message also waste the storage space on server side [10].

Twitter released its rules on eliminating and controlling spams and abuse which include the regulation on banning the spam accounts permanently, and the definition of spam accounts [11]. Unfortunately, some messages from spam accounts have been changed correspondingly to defeat these rules. Previously, researchers tried to detect spam account by building a URL blacklist. However, 90% of users have already clicked on a spam URL before it is put into blacklist [8]. Furthermore, the URL shorten service also lead to the uncertainty on detecting spam URL [12]. Currently, three major trends for detecting spam accounts include data compression algorithms [13], machine learning [14], and statistics [15]. In recent years, more and more new spam detection methods are created based on these three bases. Lin and Huang [10] evaluated the tweets by examining the URL rate and the interaction rate. Lumezanu and Feamster [5] categorized the tweets by characterizing the publishing behavior and analyzing the effectiveness of spam. Bayesian statistic was utilized for classifying the tweets [16]. It is worth to learn that most of the spam tweets are not sent by human but bots. However, bots does not only send out spam tweets, but also send out useful information sometimes, such as weather information, traffic update, and earthquake events.

The existence of bots may not be widely recognized by all the Twitter users. According to Japan Times, the bot is defined as follow: "*Twitter-bots are small software programs that are designed to mimic human tweets. Anyone can create bots, though it usually requires programming knowledge. Some bots reply to other users when they detect specific keywords. Others may randomly tweet preset phrases such as proverbs. Or if the bot is designed to emulate a popular person (celebrity, historic icon, anime character etc.) their popular phrases will be tweeted. Not all bots are fully machine-generated, however, and interestingly the term "bot" has also come to refer to Twitter accounts that are simply "fake" accounts. [17]*."

Meanwhile, cyborgs, a mix of humans and bots, refer to either bot-assisted humans or human-assisted bots [18]. A bot can send tweets automatically by called Twitter APIs. Sometimes, after a bot receives audience, the creator (human) may tweet through bots which led to a merged version of humans and bots – cyborgs. The emergence of bots and cyborgs should be attributed to the growing users population and open nature of Twitter [18]. Bots and cyborgs generate many tweets everyday by providing various information, including news updates, advertisements, emergency information, et al. [18]. Some researchers study how to increase bots' influence in social media [19]. On the other hand, identifying bot and cyborgs are not easy [20]. According to Chu and his team [8], they identified that 10.5% of Twitter accounts are bots while 36.2% are classified as cyborgs. The existence of bots and cyborgs bring both pros and cons. The information being tweeted, such as news, job posting, allow people with the access to latest updates. However, the spam tweets sent by bots are harmful to social media users.

Azmandian [21] introduce a two-steps procedure for eliminating the bots from the Twitter data. The two-steps are: (1) All the Twitter users whose tweets contain URLs more than 70% of the time will be identified as bots. (2) All the users who traveled for more than 120km/h will be identified as a bot. However, these previous research did not focus on the elimination of spams, bots and cyborgs in geo-tagged tweets. This paper will focus on how to identify and remove data noises and errors in geo-tagged tweets.

## 3. SYSTEM ERRORS IN THE STREAMING APIS FOR GEO-TAGGED TWEETS

Since the Twitter Streaming API provides the bounding box option to enable users to search and collect geo-tagged tweets within the bounding box area. Many people think that all collected geo-tagged tweets by the Streaming API will be restricted within the user-defined bounding box. However, we compared two case studies of Streaming API with bounding boxes for the County of San Diego and the City of Columbus during one month (November, 2015). Both cases illustrated that there are only 42.7% of tweets are contained within San Diego County and 83.8% of tweets are within the City of Columbus (Table 1).

**Table 1.** The percentage of geo-tagged tweets within the original bounding box or within the State boundary in the County of San Diego and the City of Columbus collected in one month (November, 2015).

| GeoViewer@SanDiego | November | Percentage to total Tweets | Outside Boundary |
|---|---|---|---|
| Tweets in San Diego County (SDC) | 97944 | 42.7% | 131464 |
| Tweets in Cali. State (excluding SDC) | 56382 | 24.6% | |
| Other Regions | 75082 | 32.7% | |
| Total Tweets | 229408 | | |
| GeoViewer@Columbus | November | Percentage to total Tweets | Outside Boundary |
| Tweets in Columbus City (CC) | 53291 | 83.8% | 10279 |
| Tweets in Ohio State (excluding CC) | 10043 | 15.8% | |
| Other Regions | 236 | 0.4% | |
| Total Tweets | 63570 | | |

Figure 1 illustrates the spatial distribution of geotagged tweets collected by the Twitter Streaming API using the bounding box of San Diego County during two months (October and November, 2015). Around 42.7% of the geo-tagged tweets are within the boundary of San Diego County and 57.3% of tweets are outside the original defined bounding box. We also conducted kernel density maps using geotagged tweets and found out that the major hot spots (clustered areas) are located around San Diego Downtown and two other hot spots are around the cities of Los Angeles and San Francisco. Figure 1 illustrates that the distribution of outside-bounding box tweets are mainly located in California and a few tweets are located in Mexico. There are a small amount of tweets located in South America, Europe, and Southeast Asia.

Figure 2 illustrates the spatial distribution of geotagged tweets using the bounding box of the City of Columbus in Ohio, USA during October and November, 2015. 83.8% of tweets are within the actual boundary of the City of Columbus. 15.8% of tweets are within the Ohio State. Only 0.4% of geotagged tweets are outside the Ohio State.

Comparing the two case studies, the San Diego case has much higher ratio of outside bounding box tweets. The Columbus case has better results of using bounding box to collect the mid-size city level tweets.

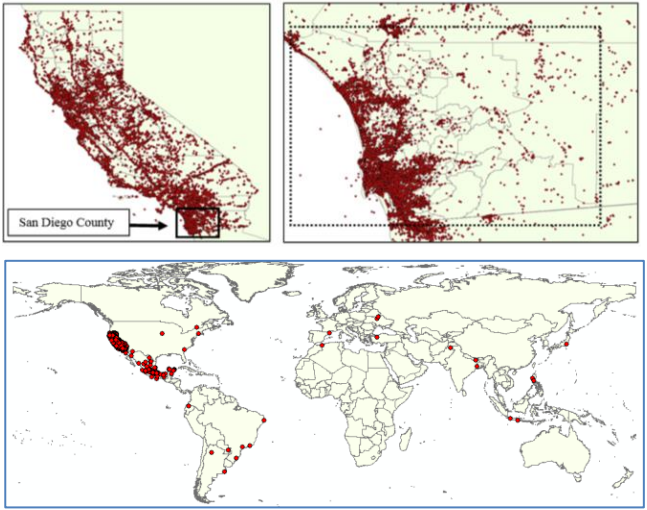

**Figure 1.** The spatial distribution of geo-tagged tweets using the Streaming API with the bounding box of San Diego County (October and November 2015) at county, state, and world scales.

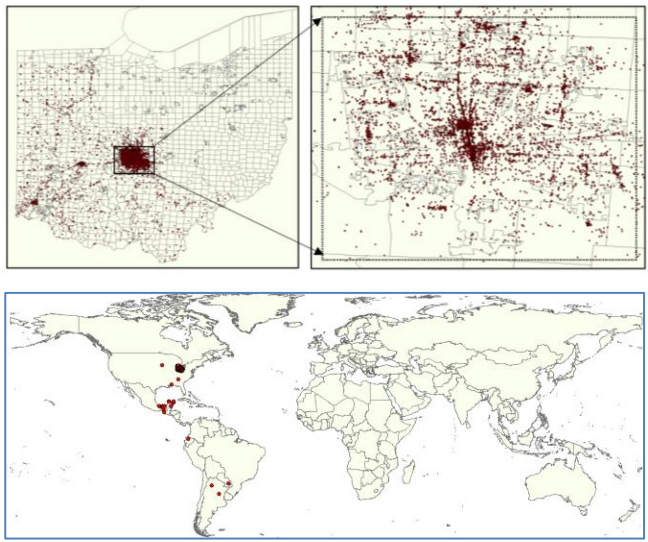

**Figure 2.** The spatial distribution of geo-tagged tweets using the streaming API with the bounding box of Columbus City (October and November 2015) at county, state, and world scales.

There is one possible explanation about this system error generated by the Twitter APIs. Based on the Twitter API documents (https://dev.twitter.com/streaming/overview/request-parameters#locations), the streaming API uses the following heuristic rules to determine whether a given Tweet falls within a bounding box:

1. *If the coordinates field is populated, the values there will be tested against the bounding box. Note that this field uses geoJSON order (longitude, latitude).*
2. *If coordinates is empty but place is populated, the region defined in place is checked for intersection against the locations bounding box. Any overlap will match.*
3. *If none of the rules listed above match, the Tweet does not match the location query. Note that the geo field is deprecated, and ignored by the streaming API.*

The Twitter rule #2 may be the key reason for collecting tweets outside the bounding box. For example, if a tweet selects "California" as the place to check-in, the place box of California will overlap with the bounding box of San Diego. Therefore, the tweet outside San Diego will be selected. However, there are still several global geo-tagged tweets which cannot be explained by the heuristic rules. We will recommend that the future users of Twitter Search APIs should add the geo-filter procedure after collecting geo-tagged tweets using bounding box methods to remove the system errors.

## 4. DATA NOISES FROM COMMERCIAL BOT AND CYBORG TWEETS

In this paper, we defined the noises in social media as the messages created by bots or cyborgs with commercial purposes (such as advertisements and marketing events) since most social media researchers focus on self-expression messages by actual human beings. After manually reviewing hundreds of geo-tagged tweets in our collection, we found out that one easy way to detect the bot or cyborg tweets for advertisement is to examine a metadata field in tweets, called "source". If a tweet was created on an iPhone device, the source field will be "Twitter for iPhone". If a tweet was created by specific bots or web programs, the source field could be "TweetMyJOBS", "dlvr.it", or "AutoCarSale". Therefore, by classifying different types of "source" values in the metadata, we can remove these noises created by bots or cyborgs for commercial purposes (Table 2).

We collected one month of geotagged tweets in San Diego County in November 2015. After removing the tweets outside the San Diego County, 97,944 tweets are within San Diego County. After reviewing hundreds of different source names in tweets, we created a black list for data noises and advertisement tweets based on the source names in San Diego (Table 2). Among the total tweets in San Diego, there are 29.42% tweets being recognized as noise data by our list. Table 3 illustrates the top six categories of noise sources, including Jobs (21.17%), Advertisement (3.49%), Weather (2.18%), Earthquake information (1.06%), News (0.97%), and Traffic (0.53%).

Similar to the previous comparison studies, we also collect one month of geotagged tweets in the City of Columbus in November 2015. Table 3 illustrated the black list for data noises and advertisement tweets in Columbus. There are 53.47% tweets being recognized as noise data in Columbus based on our black list which is much higher than the San Diego case. The top five categories of noise sources in Columbus are Jobs (43.23%), Advertisement (3.63%), Traffic (2.79%), News (2.10%), and Weather (1.72%). We cannot find the Earthquake information bots in the City of Columbus.

**Table 2.** The Proportion of data noises with different "source" names in one month of geo-tagged tweets (November, 2015) in San Diego County

| | Source category | Source name | Hashtag | Tweet number | Percentage |
|---|---|---|---|---|---|
| | Job | TweetMyJOBS | | 16005 | |
| | | SafeTweet by TweetMyJOBS | | 4726 | |
| | | CareerCenter | | 6 | |
| Total | | | | 20737 | 21.17% |
| | Advertisement | dlvr.it | | 2837 | |
| | | Golfstar | | 269 | |
| | | dine here | | 182 | |
| | | Simply Best Coupons | | 77 | |
| | | Auto City Sales | | 56 | |
| | | sp_california | Coupon | 41 | |
| Total | | | | 3421 | 3.49% |
| | Weather | Cities | | 2105 | |
| | | iembot | | 24 | |
| | | Sandaysoft Cumulus | | 7 | |
| Total | | | | 2136 | 2.18% |
| | Earthquake | | Earthquake | 762 | |
| | | everyEarthquake | | 203 | |
| | | EarthquakeTrack.com | | 69 | |
| | | QuakeSOS | | 9 | |
| Total | | | | 1043 | 1.06% |
| | News | San Diego Trends | | 843 | |
| | | WordPress.com | | 111 | |
| Total | | | | 954 | 0.97% |
| | Traffic | TTN SD traffic | | 512 | |
| | | TTN LA traffic | | 11 | |
| Total | | | | 523 | 0.53% |
| | | | Percentage of Noise: | | 29.42% |

**Table 3.** The Proportion of data noises with different "source" names in one month of geo-tagged tweets (November, 2015) in the City of Columbus

| | Source category | Source name | Hashtag | Tweet number | Percentage |
|---|---|---|---|---|---|
| | Job | TweetMyJOBS | | 16789 | |
| | | SafeTweet by TweetMyJOBS | | 6250 | |
| Total | | | | 23039 | 43.23% |
| | Advertisement | dlvr.it | | 1642 | |
| | | circlepix | | 147 | |
| | | dine here | | 77 | |
| | | Beer Menus | | 53 | |
| | | sp_ohio | Coupon | 4 | |
| | | DanceDeets | | 4 | |
| | | sp_oregon | Coupon | 3 | |
| | | SmartSearch | | 2 | |
| | | JCScoop | | 1 | |
| | | LeadingCourses.com | | 1 | |
| Total | | | | 1934 | 3.63% |
| | Traffic | TTN CMH traffic | | 1486 | |
| Total | | | | 1486 | 2.79% |
| | News | Columbus Trends | | 1021 | |
| | | eLobbyist | | 80 | |
| | | WordPress.com | | 10 | |
| | | twitterfeed | | 8 | |
| | | stolen_bike_alerter | | 1 | |
| Total | | | | 1120 | 2.10% |
| | Weather | Cities | | 578 | |
| | | iembot | | 337 | |
| Total | | | | 915 | 1.72% |
| | | | Percentage of Noise: | | 53.47% |

Figure 3 illustrates the proportion of "source" categories including both noise and non-noise tweets within one month of geo-tagged tweets (November 2015) from the Streaming API within the San Diego County boundary. The red color indicates the tweets labelled as bots or cyborgs based on our black list. The green color indicates the tweets created by generic Twitter platform (such the Android or iOS Twitter Apps). The blue color indicates the tweets from other third-party apps or services, such as Instagram or Foursquare. The most popular source (platform) in San Diego geo-tagged tweets is Instagram (46,484 tweets, 47.46%), which is a very popular social media platform for sharing photos and videos either publicly or privately on mobile devices. Instagram users can extend their photo sharing to other platforms, including Twitter and Facebook. Since the default setting of Instagram is geotag-enabled. Therefore, most Instagram messages sharing on Twitter include detail geolocations. The second popular platform is "TweetMyJOB", which contains job market advertising tweets using the geolocations of recruiting companies. Foursquare is a location-based search-and-discovery services with personalized recommendations and tips. The generic Twitter apps are the fifth for Android and the seventh for iPhone devices (Figure 3).

Figure 4 illustrates the proportion of "source" categories within the City of Columbus. The most popular platform in Columbus is the "TweetMyJOB" bots. The second popular platform is Instagram. The ranking of the rest categories are similar to the San Diego case study.

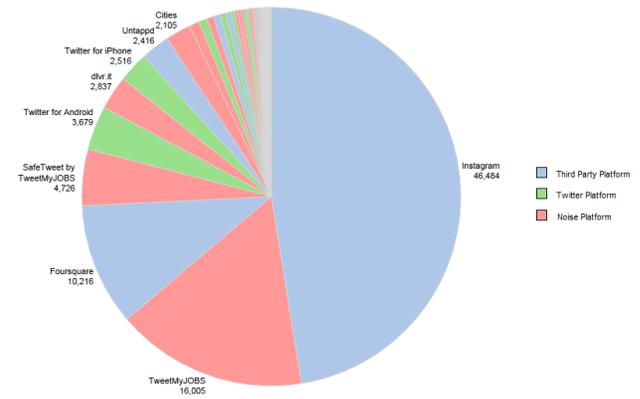

**Figure 3.** The numbers of Tweets produced by different platforms inside the San Diego bounding box during the month of November, 2015.

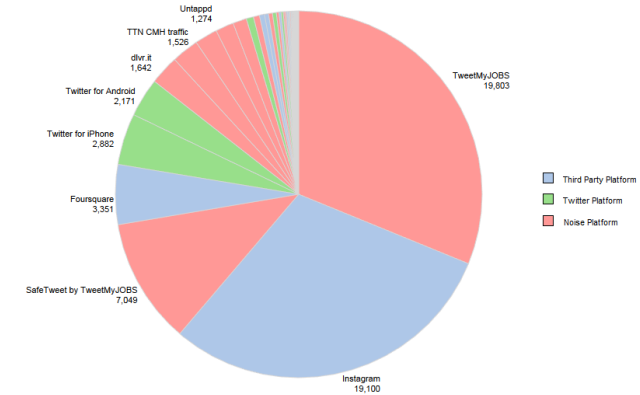

**Figure 4.** The numbers of Tweets produced by different platforms inside the City of Columbus bounding box during the month of November, 2015.

## 5. WHERE ARE THESE BOTS?

This section will focus on the visualization of the actual locations of these bot and cyborg tweets in our collected data. From a spatial analysis perspective, if we cannot remove these noises, the spatial distribution patterns of these noises may cause a significant problem or biases in the outcomes of spatiotemporal analysis.

We started to map the most popular bots in our black list: **TweetMyJOBS**, which is a Twitter-based recruiting service embedded with the job advertisement tweets. We found out that most of TweetMyJOBS tweets are around the center of cities (San Diego downtown or Columbus downtown) (Figure 5) due to the higher density of business buildings and addresses in the center of cities. The red color areas in Figure 5 are the high kernel density areas created by the clusters of tweets.

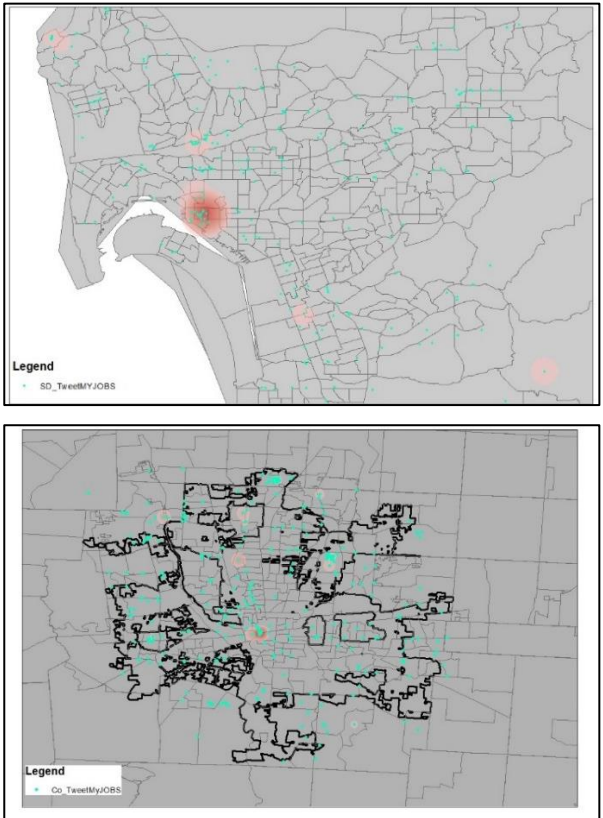

**Figure 5.** The spatial distribution of the TweetMyJOBS tweets in San Diego (top) and Columbus (bottom) during the month of November, 2015. Red color areas are the clustered tweets in these locations.

There are other spatial distribution patterns of bots or cyborgs based on their source names:

- **Cities** is one of the twitter bots for weather forecasting. Their spatial distribution is following the locations of major weather stations or the center of local neighborhood throughout the whole cities.
- **Dlvr.it** is a new service for attracting and engaging audiences across the web with powerful content sharing tools. It can help users to distribute their posted social media messages to other platforms automatically. All dlvr.it tweets are located in a single point (at the center of San Diego downtown and at the center of Columbus downtown).
- **San Diego Trends and Columbus Trends** is used for sending out the local news. There is only one location tweeting through San Diego Trends and Columbus Trends which is the locations of San Diego City Hall and Columbus City Hall.
- **TNN CMH traffic** is the platform of posting traffic accidents. The whole platform generated 1486 tweets in Columbus City during the month of November, 2015. According to the spatial distribution and kernel density map below, the spatial distribution of their tweets are highly correlated to major roads and road intersections (Figure 6).

Based on these finding, we can conclude that the spatial distributions of bots and cyborgs are not random. Different types of bots have their unique spatial distribution patterns. Researchers should remove these bot or cyborg tweets before conducting any spatiotemporal analysis using geotagged tweets.

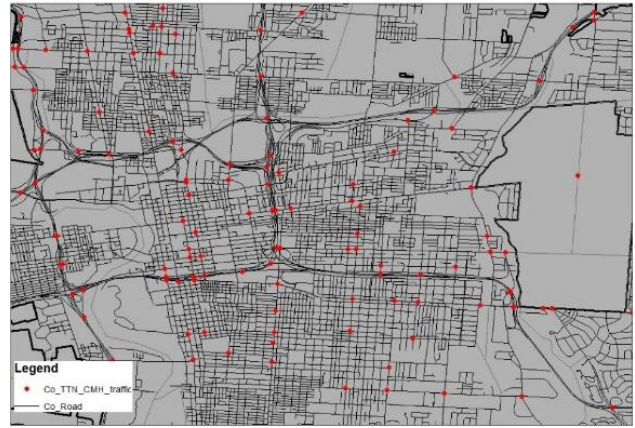

**Figure 6.** The spatial distribution of the **TNN CMH traffic** tweets in Columbus during the month of November, 2015. Red dots are the clustered tweets created by the traffic bots.

## 6. TWEETING FREQUENCY IN GEO-TAGGED TWEETS (USER BIASES)

To explore potential user biases in geo-tagged tweets, this study first calculated the frequency of geo-tagged tweets for individual users. Geo-tagged tweets were collected within the bounding boxes of San Diego County, CA and the City of Columbus, OH in U.S. throughout the month of November, 2015. After removing the bots and cyborgs, **69,317** human tweets are within San Diego bounding box and 15,916 unique Twitter users (accounts) were identified within the collected tweets. Figure 7 (left) reveals the number of users along with their geo-tagged rates throughout the whole month of November, 2015 in San Diego. Over 7,900 users only had one tweet during the whole month, which consists up to 49% of total users. More than 80% of Twitter users created less than 5 tweets in the whole month. But 1% of Twitter users created 16% of total Tweets. This finding is very similar to other literatures in Twitter message analysis [22][23].

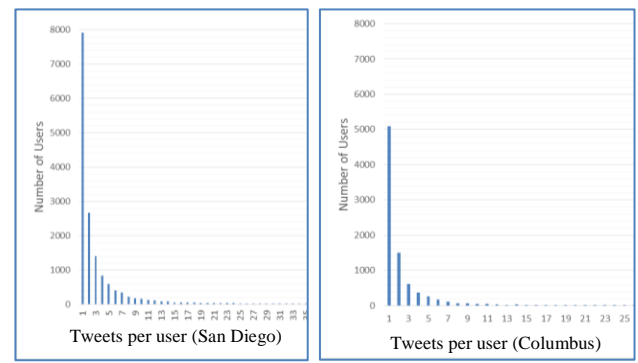

**Figure 7.** The number of users along with their geo-tagged tweeting rates within November 2015 in San Diego County (left) and in Columbus City (right).

Using the bounding box of the Columbus City, **29,902** human tweets were identified for the month of November, 2015. 8,758 unique Twitter users are identified. Over 5,000 users in Columbus only tweeted once during the whole month, which consists up to 58% of total users (Figure 7, right). Over 86% of Twitter users created less than 5 tweets in the whole month. Meanwhile, 1% of Twitter users created 19% of total Tweets.

The patterns of tweets per user in both San Diego and Columbus cases are similar (Figure 1). The San Diego case has larger number of human tweets and unique users. The Columbus case has higher portion of users with only one tweet within one month and the higher percentage of tweets created by 1% top users. Table 4 shows a side by side comparison between the two case studies.

**Table 4. User biases comparison between San Diego and Columbus cases.**

|  | Human Tweets | Human Users | Ratio of User who tweeted 1 time | 1-5 times | Most active user |
|---|---|---|---|---|---|
| San Diego | 69,317 | 15,916 | 49% | 84.20% | 903 |
| Columbus | 29,902 | 8,758 | 58% | 89.50% | 964 |

To avoid the user biases (small percentage of users creating large amounts of tweets), we can calculate the number of unique users rather than the number of tweets for statistic analysis [24]. We can also remove top 1% or 5% of active users if we need to analyze the common messages from the general population. Another possible method is to select one tweet per user for sentiment or statistic analysis.

## 7. DISCUSSION

When social media researchers and geographers utilized geo-tagged tweets for their research projects, they need to identify and remove data noises, user biases, and system errors from their collected geo-tagged tweets. This paper provides a systematic view to illustrate the characteristics of geo-tagged tweets from the Streaming API focusing their noises, user biases, and system errors. We recommend the following data process procedure for removing the bot and cyborg tweets:

1. Using the Twitter Streaming with a bounding box to collect geotagged tweets.

2. Use the same bounding box to filter out (remove) tweets outside the bounding box or the actual boundary of cities using GIS software.

3. Manually review the source fields in the collected tweets to identify top 10 or more bot or cyborg tweets and create a black list for each city.

4. Remove bot or cyborg tweets using the black list.

5. If the research needs to analyze common messages from the general population, we should use the numbers of unique users rather than the numbers of tweets in spatial analysis. Another possible solution is to remove the top 1% or 5% active users from the database.

Hopefully, by identifying these errors, biases, and noises, researchers can remove these bots and cyborgs from their raw data before conducting actual spatial analysis and improve their research findings and outcomes.

## 8. ACKNOWLEDGMENTS

This material is based upon work supported by the National Science Foundation under Grant No. 1416509, project titled "Spatiotemporal Modeling of Human Dynamics Across Social Media and Social Networks" and Grant No. 1634641, IMEE project titled "Integrated Stage-Based Evacuation with Social Perception Analysis and Dynamic Population Estimation". Any opinions, findings, and conclusions or recommendations expressed in this material are those of the author and do not necessarily reflect the views of the National Science Foundation. The authors thank other HDMA team members' contribution to this research.